\documentclass[aps,prl,%
twocolumn, 10pt,%
amsmath,%
amssymb]{revtex4-1}

\pdfoutput=1
\usepackage{graphicx}% Include figure files
\usepackage[caption=false,labelformat=parens]{subfig}
\usepackage{epsfig}
\usepackage{dcolumn}% Align table columns on decimal point
\usepackage{bm}% bold math
\DeclareMathOperator{\sgn}{sgn}
\usepackage[bookmarks=false]{hyperref}
\hypersetup{pdfauthor={Nicholas Bevins},pdfborder=0 0 0,colorlinks=true,citecolor=black,linkcolor=black,urlcolor=blue,}

\begin{document}

%\preprint{APS/123-QED}

\title{Anomalous scaling law for noise variance and spatial resolution in differential phase contrast computed tomography}

\author{Guang-Hong Chen}%
 \altaffiliation[Also at ]{Department of Radiology, University of Wisconsin-Madison}%, 600 Highland Avenue, Madison, WI 53792-1590}%
 \email{gchen7@wisc.edu}
\author{Joseph Zambelli}%
\author{Zhihua Qi}%
\author{Nicholas Bevins}%
\affiliation{Department of Medical Physics, University of Wisconsin-Madison, Madison, Wisconsin 53705, USA}

\date{\today}% It is always \today, today,

\begin{abstract}
In conventional absorption based x-ray computed tomography (CT), the noise variance in reconstructed CT images scales with spatial resolution following an inverse cubic relationship. Without reconstruction, in x-ray absorption radiography, the noise variance scales as an inverse square with spatial resolution. In this letter we report that while the inverse square relationship holds for differential phase contrast projection imaging, there exists an anomalous scaling law in differential phase contrast CT, where the noise variance scales with spatial resolution following an inverse linear relationship. The anomalous scaling law is theoretically derived and subsequently validated with phantom results from an experimental Talbot-Lau interferometer system. 
\end{abstract}

%\pacs{87.57.Q, 07.85.Fv, 87.57.cm}% PACS, the Physics and Astronomy
\maketitle

%\section{Introduction}
%
Differential phase contrast (DPC) imaging and the extension to computed tomography (DPC-CT) have attracted recent interest due to their successful implementation using low brilliance sources \cite{Weitkamp2005,Pfeiffer2006,Pfeiffer2008,Pfeiffer2007a,Huang2009,Momose2009,Qi2010,Zambelli2010}. Initial experimental results demonstrate that DPC-CT imaging may have the potential to quantitatively measure the composition of a material \cite{Qi2010} with superior contrast-to-noise ratio (CNR) when compared to conventional absorption CT imaging \cite{Zambelli2010, Herzen2009}. In order to address whether the sensitivity of the DPC-CT measurements is sufficient for a specific imaging task, two relationships must be studied: the noise variance dependence on exposure level and the noise variance dependence on spatial resolution. The determination of first relationship is critical if DPC-CT is to be considered for biomedical applications, where the minimization of radiation dose is paramount. The second relationship, noise variance vs. spatial resolution, will determine whether sufficient CNR can be generated at high spatial resolutions and acceptable dose levels. As the spatial resolution is increased to improve fine-object visualization, the noise variance increases, generally following an inverse-power law. The power is different for projection and tomographic imaging in absorption x-ray imaging, and has not been fully investigated in the case of DPC and DPC-CT. After determining the above two relationships, the potential advantages of DPC and DPC-CT over conventional x-ray imaging will become clear, allowing for proper selection of applications.

Recently, the first relationship has been addressed for both DPC projection imaging \cite{Yashiro2008} and DPC-CT \cite{Zambelli2010}. It was demonstrated that the noise variance in both DPC and DPC-CT imaging is inversely proportional to radiation dose, which is similar to the relationship in conventional absorption x-ray imaging \cite{Macovski1983,Kalender2005}, even though the physical mechanism of image formation is dramatically different. 

In this letter, we report an anomalous scaling law, where the noise variance is inversely proportional to the spatial resolution in DPC-CT. This is in stark contrast to conventional absorption CT, where noise variance is inversely proportional to the third power of spatial resolution \cite{Kak1988}. This anomalous behavior indicates that DPC-CT imaging may enable higher spatial resolution than absorption CT at the same noise variance. In the following, we will present a theoretical analysis of the relationship between noise variance and spatial resolution. Experimental phantom results are then presented to validate the theoretical analysis.

%\section{Basic imaging principles of DPC-CT}
%
In order to study the noise variance-spatial resolution relationship, a Talbot-Lau interferometer DPC and DPC-CT data acquisition setup is used. In this setup, a partially coherent x-ray beam is diffracted by a phase grating with a $\pi$-phase shift at the mean beam energy. There will be a self-image formed at the fractional Talbot distance \cite{Weitkamp2005,Suleski1997,Arrizon1995},
\begin{equation}
d=\frac{2m-1}{16}Z_T,
\end{equation}
where $m=1,2,3,\ldots$, and $Z_T={2p^2}/{\lambda}$ is the Talbot distance, determined by the wavelength $\lambda$ and the pitch $p$ of the phase grating. In order to record the diffracted beam intensity modulations, a homodyne technique is introduced at the fractional Talbot distance \cite{Kafri1990}. An absorption grating with the same period as the diffracted beam modulation pattern ($p_2=p/2$) is used to analyze the beam modulation. After the homodyne analysis, the intensity profile is both low-pass filtered and recorded by the detector elements. At each detector element $(x,y)$, the recorded intensity can be written as \cite{Momose2003,Weitkamp2005,Yashiro2008,Pfeiffer2006,Pfeiffer2008,Pfeiffer2007a}
\begin{equation}
\label{IntensityProfile}
I(x,y)=I_0+I_1\cos\left[\frac{2\pi}{p_2}x+\phi(x,y)\right],
\end{equation}      
where phase shift in the image intensity $\phi(x,y)$ is related to the phase change $\Phi$ of the x-ray wave induced by the image object by
\begin{equation}
\label{refractionAngle}
\phi(x,y)=\frac{\lambda d}{p_2}\frac{\partial\Phi}{\partial x}=-\frac{2\pi d}{p_2}\frac{\partial}{\partial x}\int dz\delta(x,y),
\end{equation} 
where $\delta$ is the decrement of the refractive index $n=1-\delta+i\beta$. Therefore, once phase shifts of the beam intensity profiles are measured from different view angles around the image object, an image of the local distribution of the refractive index decrement can be reconstructed \cite{Faris1988,Pfeiffer2008,Qi2008}. In this letter, $\phi(x,y)$ is referred to as the projection data and Eq. (\ref{refractionAngle}) is referred as the fundamental imaging equation of DPC-CT, connecting a measurable quantity, $\phi(x,y)$, to a line integral.

%\subsection{Noise model of DPC projection data} 
%
In order to measure the projection data $\phi(x,y)$, a phase-stepping method \cite{Momose2003,Weitkamp2005} is used in which the analyzer grating is translated by a fraction of the grating pitch along the $x_g$ axis: $x_g=jp_2/M, (j=1,2,...,M)$. The measured beam intensity at each phase step is given by
\begin{equation}
\label{Phase-Stepping}
I^{(j)}(x,y)=I_0+I_1\cos\left[\frac{2\pi}{M}j+\phi(x,y)\right].
\end{equation}  
By multiplying both sides of Eq. (\ref{Phase-Stepping}) by $\exp(-ij2\pi/M)$ and summing over $j$, one obtains
\begin{equation}
\label{phase-1}
e^{i\phi(x,y)}=\frac{2}{MI_1}\sum_{j=1}^{M}I^{(j)}(x,y)\exp\left(-ij\frac{2\pi}{M}\right),
\end{equation}  
or equivalently, 
\begin{equation}
\label{phi-xy}
\tan\left[\phi(x,y)\right]=-\frac{\sum_{j=1}^{M}I^{(j)}\sin(2\pi j/M)}{\sum_{j=1}^{M}I^{(j)}\cos(2\pi j/M)}.
\end{equation}
Namely, after a Fourier transform of the measured intensity profile over all phase steps, one can obtain the desired projection data.

Due to photon number fluctuations, the projection data will fluctuate about a mean value $\bar{\phi}$, i.e., $\phi=\bar{\phi}\pm\Delta\phi$. Using the standard error propagation formula, one can calculate that the noise variance $\sigma_{\phi}^2$ of the projection data is determined by the noise variance $\sigma_{I^{(j)}}^2=\bar{I}^{(j)}$ of the measured beam intensity at each phase step:
\begin{equation}
\label{sigma-square}
\sigma_{\phi}^2=\frac{2}{\epsilon^2}\times\frac{1}{M\bar{I}_0},
\end{equation}
where 
\begin{equation}
\label{efficiency}
\epsilon=\frac{\bar{I}_1}{\bar{I}_0}.
\end{equation}
Here the parameter $\epsilon$ describes the effective efficiency of the interferometer \footnote{Note that we have ignored the small angle scattering contribution here. If small angle scattering is included, the visibility of the fringe pattern will be reduced.}. From Eq. (\ref{sigma-square}) one can see that the noise variance of the projection data is inversely proportional to the total mean exposure level $M\bar{I}_0$ at a given projection view angle. It is also inversely proportional to the square of the efficiency of the interferometer, meaning that higher interferometer efficiency will result in lower noise variance in the projection data.

When an image object rotates, the projection data are measured from different directions, resulting in data at different view angles being uncorrelated. As a result, the image noise in the projection data has a white-noise-like behavior:
\begin{equation}
\label{white-noise}
\overline{\Delta\phi_i\Delta\phi_j}=\sigma_{\phi}^2\delta_{ij},
\end{equation}
where $\delta_{ij}=1$ for $i=j$ and $\delta_{ij}=0$ for $i\neq j$, and $\overline{\Delta\phi_i\Delta\phi_j}$ is the correlation of the projection data with respect to view angle. The indices $i,j$ denote the view angle index in the DPC-CT data acquisition. This property is similar to conventional absorption CT, where the projection data also demonstrates a white noise behavior \cite{Macovski1983}.

%\subsection{Spatial resolution dependence of noise variance in DPC-CT}
%
Using the derived noise model for DPC-CT, we can analyze how the noise variance is related to spatial resolution in DPC-CT imaging. DPC-CT images can be directly reconstructed using filtered backprojection with a Hilbert filtering kernel \cite{Pfeiffer2007a,Faris1988,Huang2006,Qi2008}:
\begin{equation}
\label{recon-formula}
\delta(x,y)=\int_{0}^{\pi}d\theta F(\theta,x\cos\theta+y\sin\theta),\\
\end{equation}
where the filtering step $F$ is defined by
\begin{equation}
\label{filtration}
F(\theta,z)=\frac{p_2}{2\pi d}\int_{-\frac{\omega_N}{2}}^{+\frac{\omega_N}{2}}d\omega\left[\frac{\sgn(\omega)}{2\pi i} \tilde{\phi} (\omega,\theta)\right]e^{i2\pi\omega z},
\end{equation} 
where $\omega_N=1/(\Delta x)$ is the bandwidth of Nyquist frequency, determined by the target spatial resolution $\Delta x$ of the reconstructed images. In Eq. (\ref{filtration}), $\tilde{\phi}(\omega,\theta)$ is the Fourier transform of the projection data $\phi(x,y)$ at view angle $\theta$. Based on this definition, the noise variance of the reconstructed image at image pixel $(x,y)$ is
\begin{eqnarray}
\label{definition-noise-pixel}
\sigma_{\delta}^{2}(x,y)&=&\left(\frac{p_2}{4\pi^2 d}\right)^2\int\int d\theta_1d\theta_2\int\int d\omega_1 d\omega_2 \nonumber\\
&& \times \sgn(\omega_1)\sgn(\omega_2)\overline{\Delta\tilde{\phi} (\omega_1,\theta_1)\Delta\tilde{\phi} (\omega_2,\theta_2)} \nonumber\\
&& \times e^{i2\pi x(\omega_1\cos\theta_1-\omega_2\cos\theta_2)} \nonumber\\
&& \times e^{i2\pi y(\omega_1\sin\theta_1-\omega_2\sin\theta_2)}.
\end{eqnarray}
Using Eq. (\ref{white-noise}), one can demonstrate that the Fourier transform of the white noise is given by
\begin{equation}
\label{noise-fourier}
\overline{\Delta\tilde{\phi} (\omega_1,\theta_1)\Delta\tilde{\phi} (\omega_2,\theta_2)}=\frac{\pi D\sigma_{\phi}^2}{N}\delta(\theta_1-\theta_2)\delta(\omega_1-\omega_2),
\end{equation}
where $D$ is the detector element width used for normalization purposes, and $N$ is the total number of view angles.  Using the above relationship, Eq. (\ref{definition-noise-pixel}) can be simplified to
\begin{eqnarray}
\label{noise-pixel}
\sigma_{\delta}^{2}(x,y)=&&\frac{D\sigma_{\phi}^2}{16\pi^2 N}\left(\frac{p_2}{d}\right)^2\int_{-\frac{\omega_N}{2}}^{+\frac{\omega_N}{2}} 1 d\omega \nonumber\\
=&&\left(\frac{p_2}{4\pi d}\right)^2\frac{\sigma_{\phi}^2 D}{N\Delta x},
\end{eqnarray}
where $\sigma_{\phi}^2$ is given by Eq. (\ref{sigma-square}). Equation (\ref{noise-pixel}) says that the noise variance of the DPC-CT image is pixel independent and is inversely proportional to the spatial resolution of the reconstruction.

%\section{Experimental  methods and results}
%
In order to validate the theoretical analysis, we use an experimental Talbot-Lau interferometer system constructed at the University of Wisconsin-Madison. The data acquisition system consists of three x-ray gratings, a rotating-anode x-ray tube (G1592, Varian Medical Systems, California, USA) with a 0.3 mm nominal focal spot connected to a generator (Indico 100, CPI, Ontario, Canada), a CMOS flat panel x-ray detector (Rad-icon, Shad-o-Box 2048, California, USA) with 48 $\mu$m primitive detector pitch across a $1024\times 2048$ array, and a rotating motion stage to enable tomographic acquisitions. The three x-ray gratings were fabricated at the University of Wisconsin-Madison using similar techniques as described in literature \cite{david2007}. The phase grating was designed to introduce the differential $\pi$-phase shift with a 50\% duty cycle at a mean beam energy of 28 keV. The analyzer grating has a pitch $p_2$ of 4.5 $\mu$m. In this work, 8 phase steps were used, sampled over the 4.5 $\mu$m period. 

\begin{figure}%
\subfloat[]{\label{subfig:hcPhantom}\includegraphics[width=0.4\columnwidth]{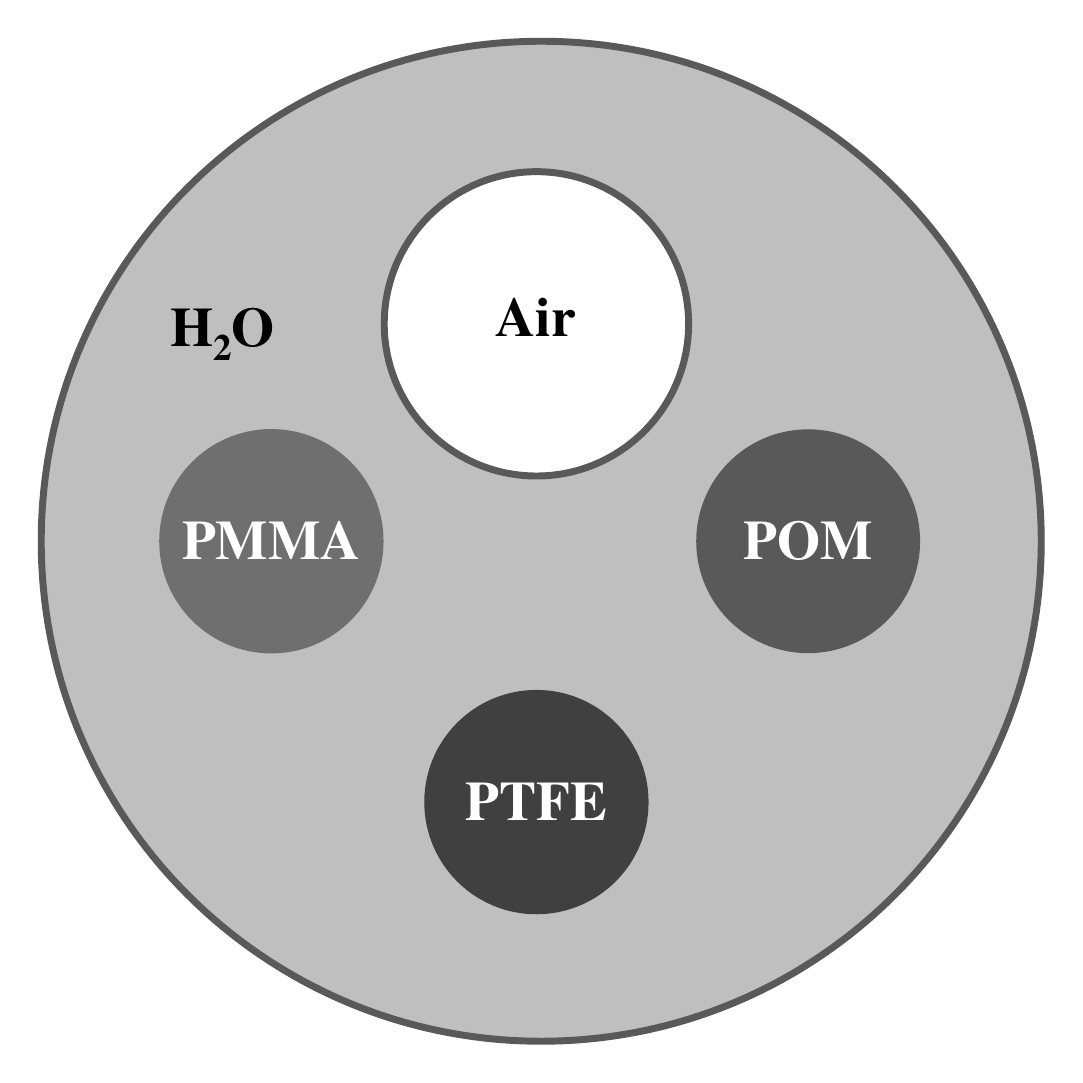}}\qquad
\subfloat[]{\label{subfig:recon}\includegraphics[width=0.4\columnwidth]{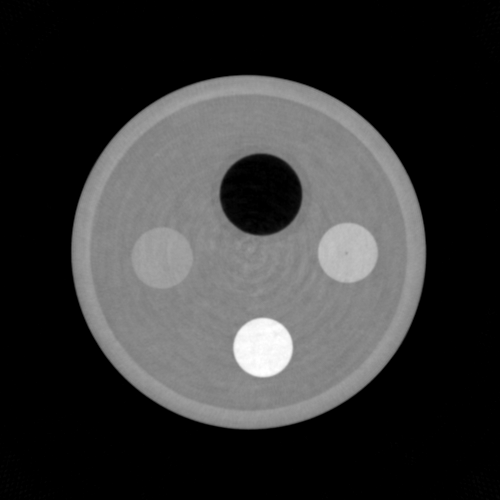}}
\caption{Cross-sectional view  and reconstruction of the imaging phantom. (a) Three plastic rods and an air-filled tube are inserted into a water-filled cylinder. (b) Reconstructed DPC-CT slice from the $2\times2$ binning data set with (80 $\mu$m)$^3$ voxels.}%
\label{fig:phantom}%
\end{figure}

The noise properties were measured in a water-filled phantom chamber with an inner diameter of 25.4 mm and wall thickness of 1.55 mm. The phantom also contained 4.76 mm diameter PTFE, PMMA, and POM rods, as well as an air-filled tube with an inner diameter of 6.60 mm and wall thickness of 0.80 mm. A cross-sectional view is shown in Fig. \ref{fig:phantom}(a).

To acquire a complete data set for CT reconstruction, 360 views of projection data were taken at $1^{\circ}$ increments. Each projection had a total exposure time of 40 seconds, divided over 8 phase steps. The tube potential was 40 kVp, with a continuous tube current of 20 mA. Once the intensity modulation was recorded, the data was processed to extract the differential phase and absorption projections.

Prior to reconstruction, the detector pixels were binned $1\times 1$, $2\times 2$, $3\times 3$, $4\times 4$, and $5\times 5$ to explore noise variance dependence on spatial resolution. The DPC-CT images (Fig. 1(b)) were reconstructed using the FBP algorithm outlined in Eq. (\ref{recon-formula}) and (\ref{filtration}), while the absorption CT images were reconstructed using a standard FBP algorithm \cite{Kak1988}. Two identical scans were performed and the reconstructions were subtracted to obtain a noise-only image. The subtracted image was divided by $\sqrt{2}$ to account for the additive noise incurred from the subtraction of two independent volumes. Due to a geometrical magnification factor of 1.2, the reconstructed voxel dimensions range from (40~$\mu$m)$^3$ to (200 $\mu$m)$^3$.

\begin{figure}%
\subfloat[]{\label{subfig:diffNoise}\includegraphics[width=.9\columnwidth]{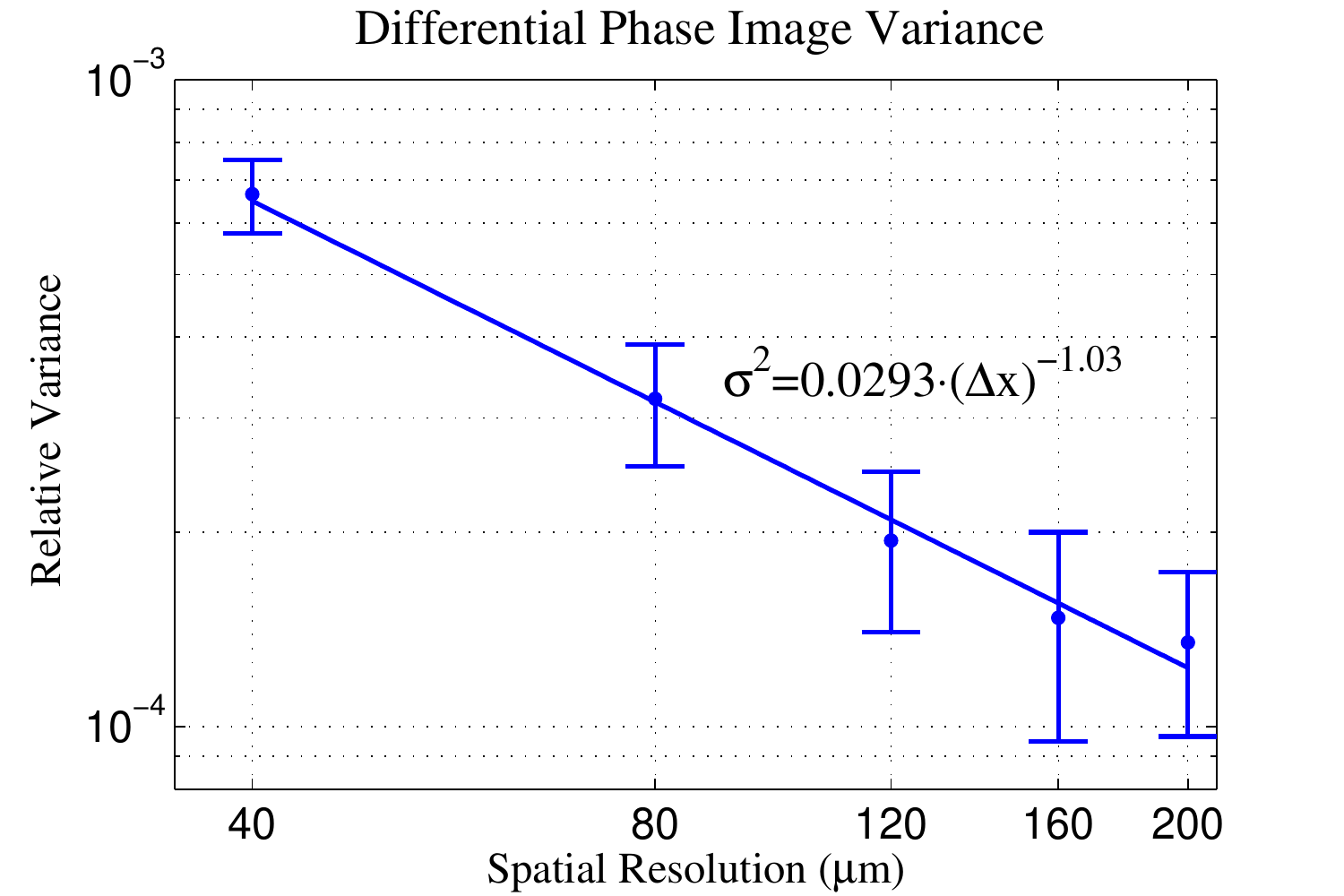}}\\
\subfloat[]{\label{subfig:attenNoise}\includegraphics[width=.9\columnwidth]{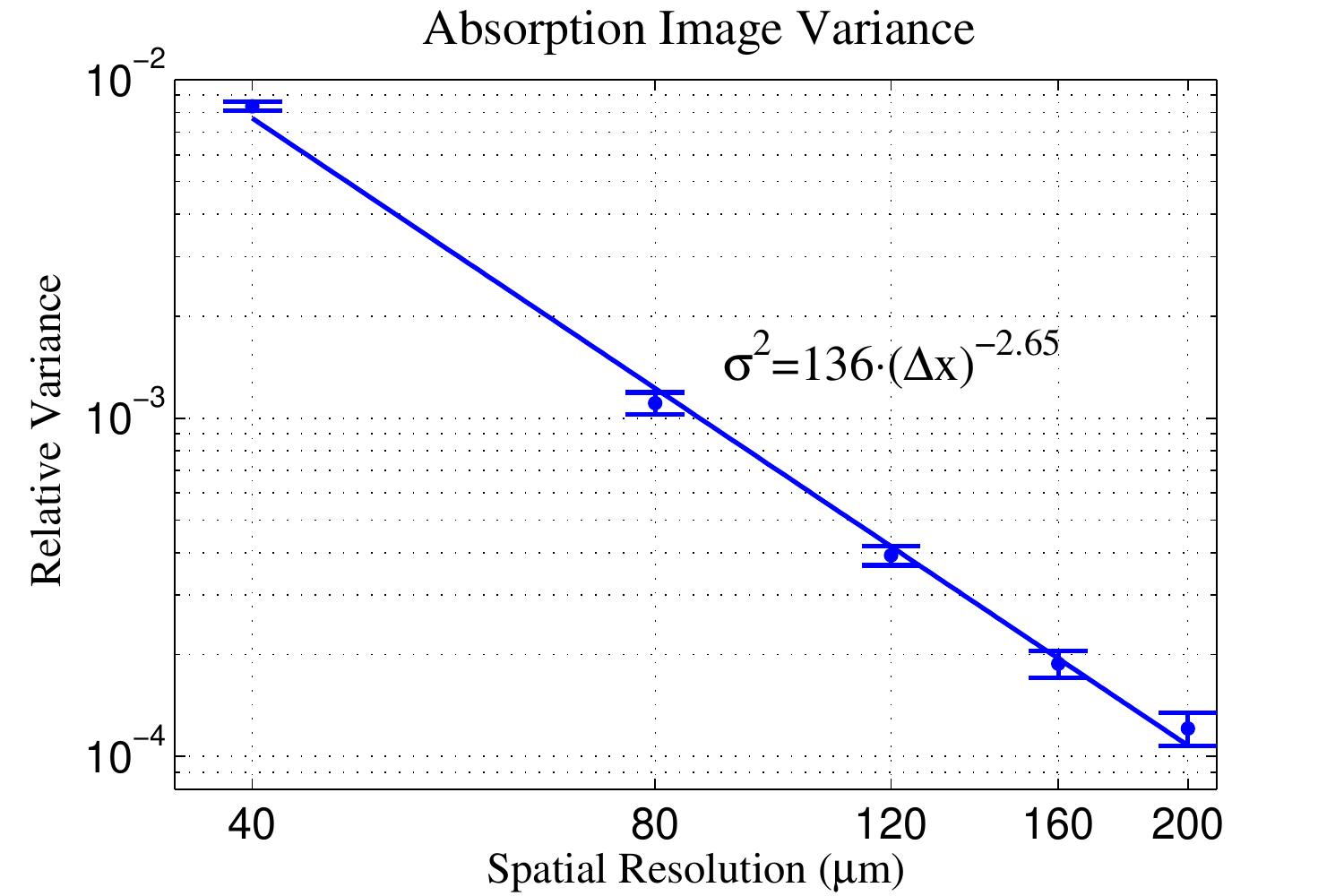}}
\caption{Log-log plots of relative noise variance against spatial resolution. (a) shows the DPC-CT results, where the fitted curve is of the form $\sigma^2\propto\Delta x^{-1.03}$, with $R^2=0.990$. (b) shows the absorption CT results, where the fitted curve is of the form $\sigma^2\propto\Delta x^{-2.65}$, with $R^2=0.997$. Note that the use of a log-log plot results in a linear display of the data, where the exponent of the fit becomes the slope of the curve.}
{\label{fig:results}}
\end{figure}

Figure \ref{fig:results} presents experimental data to demonstrate the relative variance dependence on spatial resolution. In order to measure the variance, the standard deviation was measured in the water background and then divided by the mean signal value to determine the relative noise, and the relative noise was then squared. The relative noise was used to calculate the variance because division by the mean signal level makes it a dimensionless quantity, allowing for comparison across contrast mechanisms. The logarithm of the measured data is calculated and fit to a linear function against the logarithm of spatial resolution using a least-squares method. The slope of the resulting fit is equivalent to the exponential dependence of the measured data on spatial resolution. As shown in Fig. \ref{fig:results}(a), for DPC-CT, the noise variance increases with the increase of spatial resolution as $\sigma^2_{\rm{PCCT}}\propto (\Delta x)^{-1.03}$, indicating that the noise variance is inversely proportional to the spatial resolution, as derived in Eq. (\ref{noise-pixel}). In contrast, for absorption CT [Fig. \ref{fig:results}(b)], the noise variance changes with spatial resolution as $\sigma^2_{\rm{ACT}}\propto (\Delta x)^{-2.65}$.

%\section{Discussion}
%
In the theoretical derivation, we demonstrated that the noise variance in DPC is inversely proportional to photon number. Because the number of photons is proportional to the area of the detector elements at a fixed photon flux, the number of photons scales as the square of the detector element dimension. Thus, the noise variance of the DPC projection data is proportional to the inverse square of detector pitch. It is well known that the same property is found in absorption projection imaging \cite{Macovski1983}. Therefore, for projection data, there is no difference between absorption and DPC imaging. However, as the objective of a tomographic reconstruction is to restore the depth information which is lost in projection data due to the line integral along the depth direction, one may intuitively expect that the noise variance will have an additional dimension along the depth direction. This naturally results in an inverse cubic dependence on the spatial resolution in absorption CT. When the same intuitive argument is applied to DPC-CT, one would expect that the same cubic power will appear in DPC-CT, due to the fact that both DPC and absorption CT projection data have the same noise-spatial resolution dependence. Surprisingly, this is not the case, as was shown in the theoretical and experimental results. The noise properties of the reconstructions from the two contrast mechanisms differ due to the difference in reconstruction algorithms. In absorption CT, the ramp filter amplifies the high spatial frequency content, thus amplifying the noise. However, in DPC-CT, the filtering kernel is a Hilbert kernel, which equally weights all spatial frequency content. As a result, it does not result in an amplification of the high spatial frequency noise content. 

Because the main conclusion of the paper is intrinsically determined by the filtering kernel of the reconstruction algorithm, the same conclusion of the relationship between noise and spatial resolution can be drawn for diffraction enhanced CT imaging \cite{Huang2006} and neutron imaging \cite{Pfeiffer2006a}, as each also measures refraction angle data. The data is similarly related to the derivative of a line integral as in Eq. (\ref{refractionAngle}).

%\section{Conclusion}
%
In conclusion, we have theoretically predicted and experimentally validated a novel scaling relationship between noise variance and spatial resolution. This scaling law dictates that the noise penalty significantly drops at high spatial resolution for DPC-CT. For human visualization, a minimal amount of CNR is often fixed to be between three and five \cite{Rose1973}. Therefore, because there is a smaller noise penalty at high spatial resolution, a smaller radiation dose is needed to maintain sufficient CNR for visualization. This is an advantage over conventional absorption CT, where the dose penalty often hinders the application of CT in high spatial resolution imaging.

\bibliography{PCTnoise}

\end{document}